\documentclass[twocolumn,nofootinbib]{revtex4-2}

\usepackage{amsmath}
\usepackage{amsfonts}
\usepackage{amssymb}
\usepackage{graphicx}
\usepackage{pifont}

\usepackage{color}
\usepackage{verbatim}

\usepackage{simplemargins}
\settopmargin{1.8cm}
\setbottommargin{2.4cm}

\newcommand{\D}{\mathrm{d}}

\begin{document}

% Be sure to use the \title, \author, \affiliation, and \abstract macros
% to format your title page.  Don't use lower-level macros to  manually
% adjust the fonts and centering.

\title{Lorentz contraction of electric field lines for a point charge in uniform motion}
%\title{A note on dynamic electric fields}

\author{Petar \v{Z}ugec}
\email{pzugec@phy.hr, pzugec.phy@pmf.hr}
\affiliation{Department of Physics, Faculty of Science, University of Zagreb, Zagreb, Croatia}

\author{Davor Horvati\'{c}}
\affiliation{Department of Physics, Faculty of Science, University of Zagreb, Zagreb, Croatia}

\author{Ivica Smoli\'{c}}
\affiliation{Department of Physics, Faculty of Science, University of Zagreb, Zagreb, Croatia}

%\author{xxx}
%\affiliation{x}

%\date{\today}

\begin{abstract}

We examine a logical foundation of depicting a Lorentz contraction of a Coulomb field (an electric field of a point charge in uniform motion) by means of the `Lorentz contracted' field lines. Two existing arguments for a contraction of field lines sound appealing and lead to very simple calculations yielding the correct results. However, one of them is a victim to subtle logical weaknesses, as it relies on ascribing a degree of physical reality to the electric field lines. The other one correctly proves what it sets out to prove. But it  does not provide a proof, or even a suggestion, of an additional result that can be obtained by a new poof that we present here. Though our idea is very simple, the calculations used to prove it---based on a little known, half a century old result by Tsien---are somewhat more involved than those from past arguments. 

\end{abstract}

\maketitle

\section{Introduction}
\label{intro}

Depicting vector fields by means of field lines is a widespread and time honored practice in pedagogical expositions of electric and magnetic fields, as well as in teaching a general vector field concept. However, it needs to be understood from the start that the field lines are not real physical objects by any stretch of imagination. They form an arbitrary family of integral curves, built upon a more fundamental concept of a vector field. A single and sufficient principle for a construction of field lines is that at every point they be tangential to a corresponding vector field. However, not relying on any other input but the vector field itself, they are \textit{informationally redundant}. To every point in space (and time) a vector field assigns a value and a direction of a given vectorial quantity. This procedure exhausts all information that exists and can possibly exist about this vectorial quantity. Therefore, \textit{a posteriori} constructions such as field lines do not carry any additional information that is not already present in the vector field itself~\cite{redundant}.

Still, field lines are often very convenient and efficient way of representing some properties of a given vector field. A reason for this is simple: they do not inherit \textit{all} information from a vector field, but only the information on its direction. Thus, this reduction of information makes them more accessible to a viewer. In other words, they are `easier on the eyes and mind' than some, more formally appropriate vector field representations. An example of these appropriate but `heavy' representations are vector field plots, which are often informationally `too dense' for an effortless interpretation. For this reason (and probably some historical ones) a literature abounds with field line representations of both static and dynamic vector fields. In that, the representations of dynamic fields may be especially efficient if a representation itself is made dynamic; for example, by animating the evolution of field lines, following the evolution of a field~\cite{motion}.

However, while the field line diagrams reduce the amount of information to be processed by viewer, they also introduce some informational artifacts stemming from their definition, i.e. from a principle used for their construction (that they be tangential to a vector field). This constraint obviously has nothing to do with the nature of a vector field itself; it is just a reflection of our own preferences and `biases' toward visually continuous and connected structures. As such, this \textit{ad hoc} requirement will result in the field line properties going beyond a vector field itself. It indeed constitutes something `new' about the field lines themselves, \textit{but it does reveal anything new about a vector field}, does not add anything new to it. Moreover, this `forced' mathematical property severs them from physical reality, as the field lines are not subject to any physical law. In general, they certainly do not correspond to particle trajectories within a vector field, which seems to be one of the popular misconceptions among students~\cite{trajectory_0,trajectory_1,trajectory_2,trajectory_3,trajectory_4}. Therefore, in using field lines for teaching or learning purposes---or in purposefully analyzing a particular vector field---one must be exceedingly careful not to confuse the field line properties coming from their definition, with the properties coming from a `generating' vector field. An immediate danger in failing to do so is an introduction of various misconceptions, either about the field lines of the vector fields themselves.

Unfortunately, a literature abounds with field line misconceptions: false and/or incoherent claims not only about field lines but also about the vector field properties deduced from them. A body of literature attempting to counter these issues has also been produced, focusing mostly on the misconceptions about the electric~\cite{redundant,electric_1,electric_2,electric_3,electric_magnetic} and magnetic~\cite{redundant,electric_magnetic,magnetic_1,magnetic_2,magnetic_3,magnetic_4} field lines. To this day it has met with limited success, as witnessed by many studies into students' understanding of the field line concept and their struggles with it~\cite{trajectory_0,trajectory_1,trajectory_2,trajectory_3,trajectory_4,students_0,students_1,students_2,students_3}.

\begin{figure*}[t!]
\centering
\includegraphics[width=0.32\textwidth,keepaspectratio]{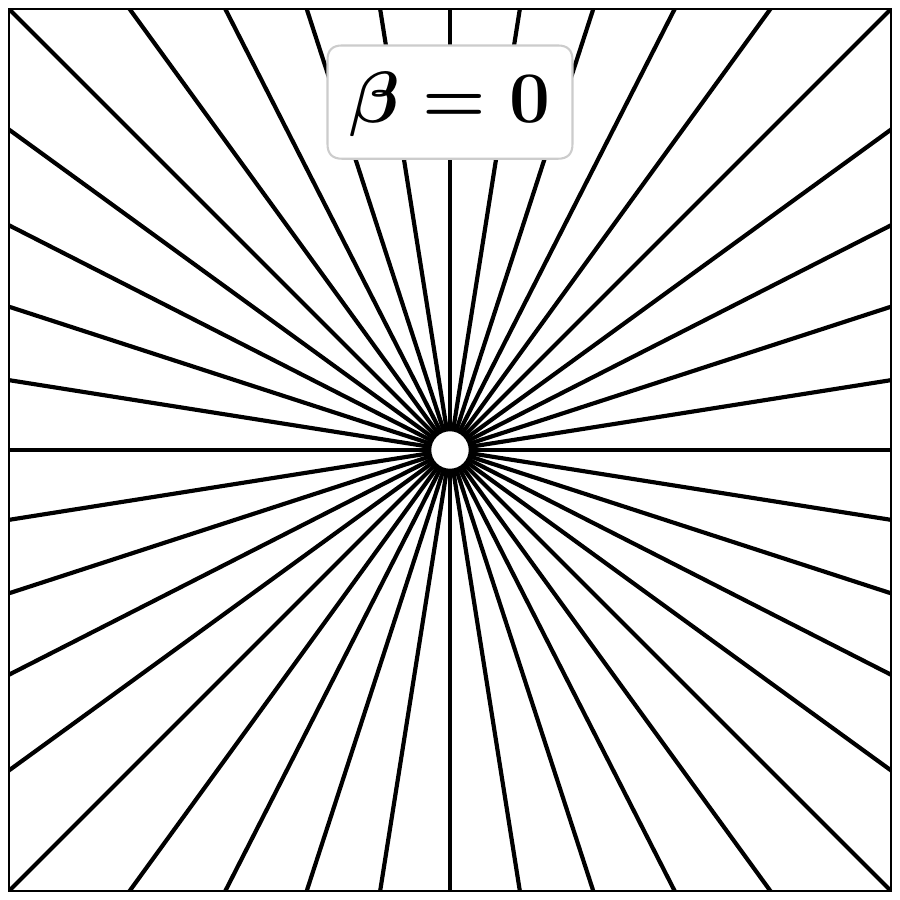}
\includegraphics[width=0.32\textwidth,keepaspectratio]{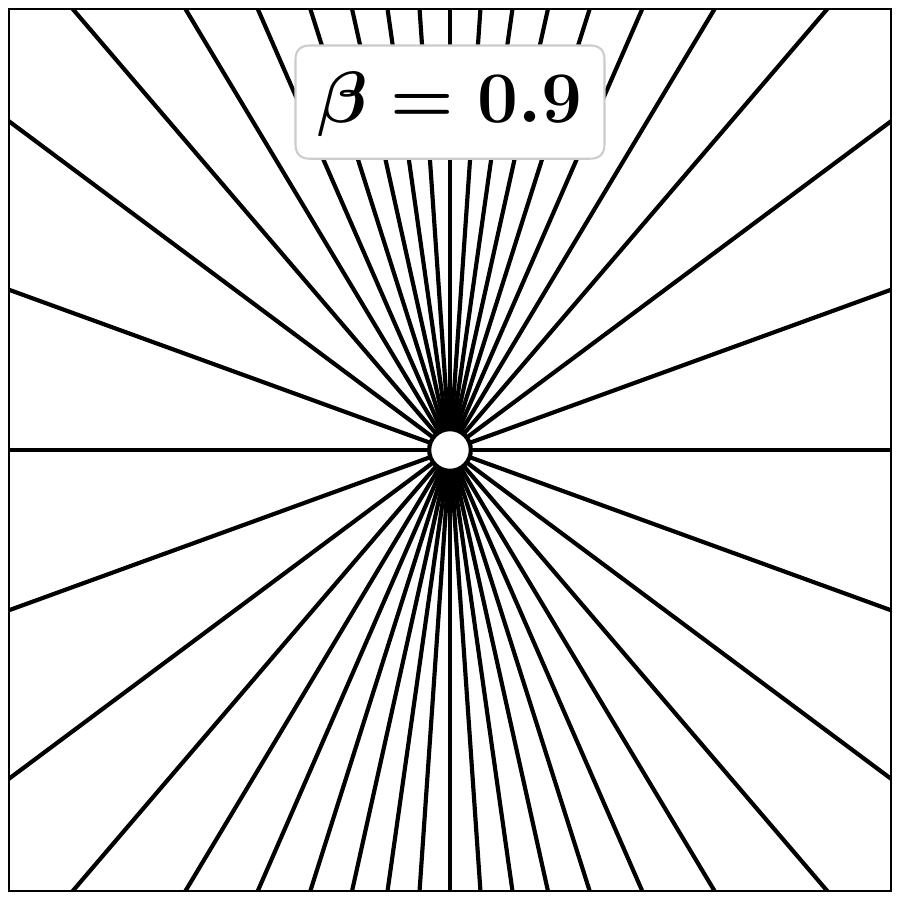}
\caption{Left: isotropic configuration of field lines depicting an electric field of a point charge at rest. Right: `Lorentz contracted' field lines depicting an electric field of a point charge in a uniform horizontal motion (either to the left or right) with \mbox{$\beta=0.9$}.} 
\label{fig1}
\end{figure*}

The main body of this work addresses a subject of representing an electric field of a point charge in uniform motion by means of the `Lorentz contracted'\footnote{
Since field lines are not physical objects, they are not subject to a true Lorentz contraction. Thus, we use the term `Lorentz contracted field lines' only for lack of a clearer phrase describing a field line transformation in question.  % more compact and self-explanatory
} field fines. Since there is no physical reality to the density of field lines, there is no \textit{a priori} reason to draw them in any specific way. It would be perfectly correct to draw them isotropically or distributed in any other way. How we draw them critically depends on the message we are trying to convey. That is precisely why the arguments for drawing them as Lorentz contracted must be well founded. In the context of conveying a message about a Lorentz contraction of Coulomb field, drawing the Lorentz contracted field lines \textit{is} justified. However, we find that common arguments for it (presented in Section~\ref{old_arg}) feature some subtle logical weaknesses or nonobvious limitations. Hence, we provide here an improved proof supporting the procedure. Our proposal should not be interpreted so much as a critique of past arguments, but rather as an improvement upon them and an additional support for their claim.

In light of a recent \textit{direct} experimental verification of a Lorentz contraction of Coulomb field~\cite{nature}, we expect a second renaissance in teaching the relativistic transformations of electric and magnetic fields. In view of various misconceptions related to a subject of field lines---a tool widely used in representing the Lorentz transformations of these fields---we feel that every bit of logical correctness about them is very much warranted. In a previous work~\cite{point_charge} we have already addressed some `low-level' confusions and less known facts about an electric field of a point charge in uniform motion. Here we continue in a similar vein, tackling a slightly more technical subject.

\section{Existing arguments for a Lorentz contraction of electric field lines}
\label{old_arg}

Electric field of a point charge at rest is usually represented by a set of isotropically distributed field lines around a charge (figure~\ref{fig1}, left plot). The idea behind such display is that an isotropy of \textit{selected} field lines represents an isotropy of a familiar electrostatic field:
\begin{equation}
\mathbf{E}(\mathbf{r})=\frac{q}{4\pi\epsilon_0}\frac{\hat{\mathbf{R}}}{R^2},
\label{static_field}
\end{equation}
with $q$ as a value of the charge, $\epsilon_0$ the permittivity of vacuum and \mbox{$\mathbf{R}=\mathbf{r}-\mathbf{r}_0$} as charge-relative position of a field observation point $\mathbf{r}$, wherein a charge itself is located at $\mathbf{r}_0$. In that: \mbox{$R=|\mathbf{R}|$} and \mbox{$\hat{\mathbf{R}}=\mathbf{R}/R$}.

\begin{figure*}[t!]
\centering
\includegraphics[width=0.32\textwidth,keepaspectratio]{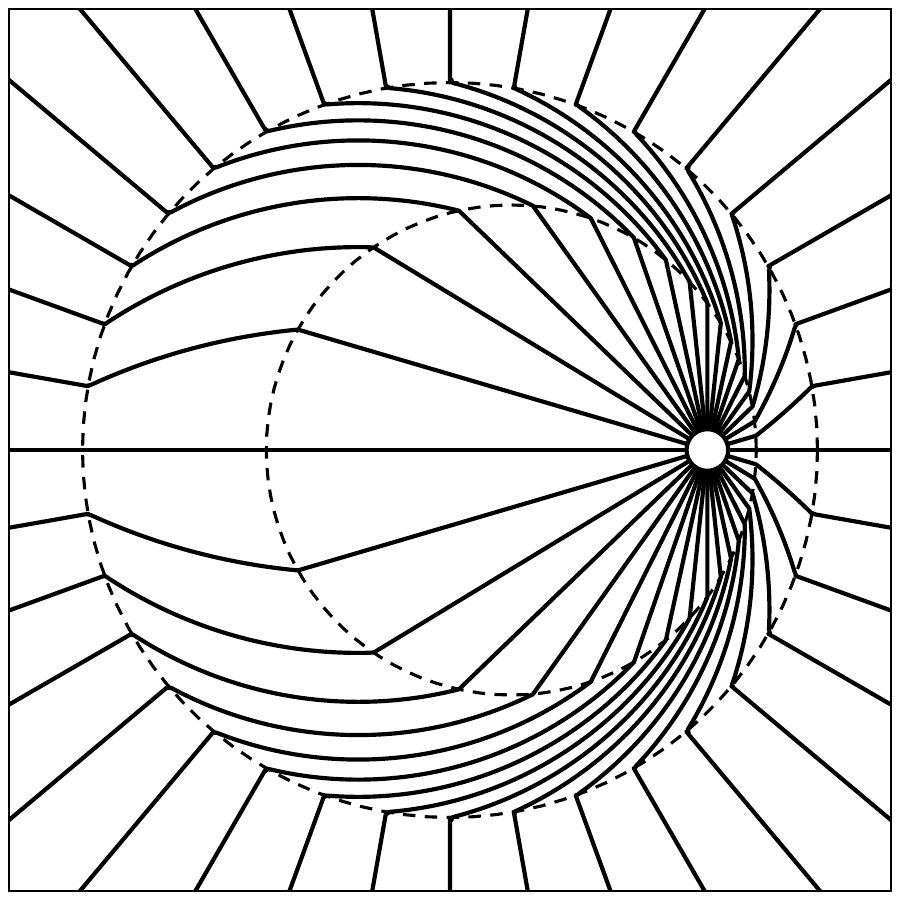}
\includegraphics[width=0.32\textwidth,keepaspectratio]{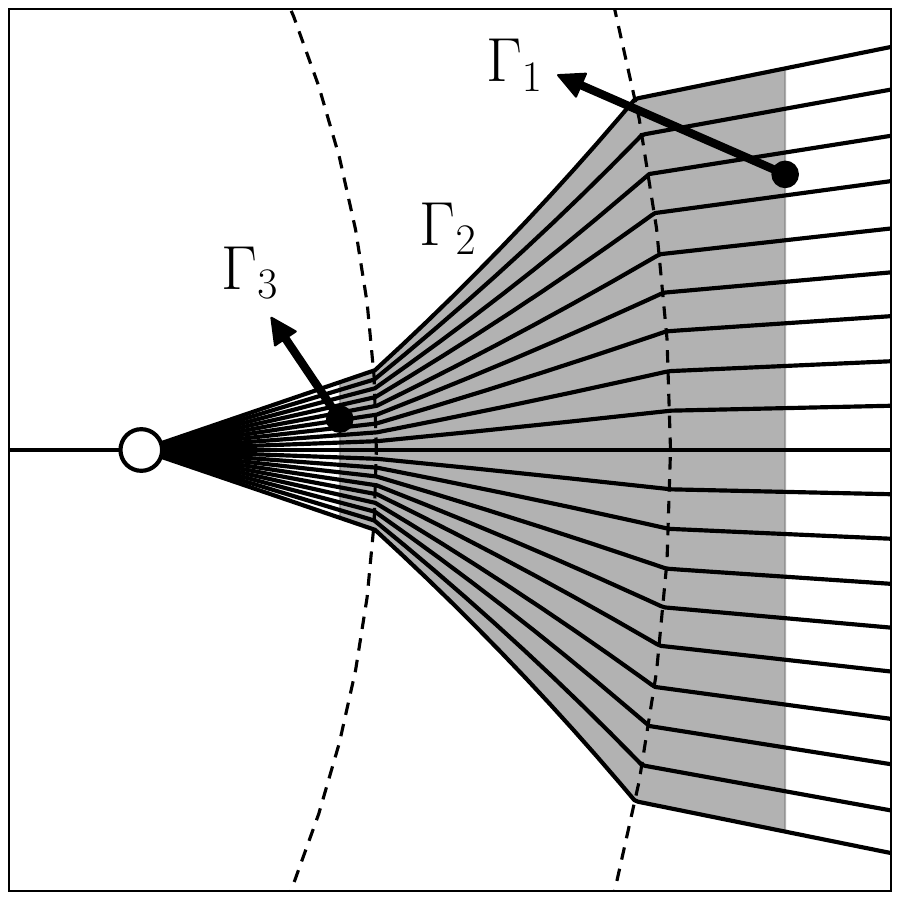}
\caption{Field lines of a point charge uniformly accelerated from rest to a state of uniform motion to the right, with \mbox{$\beta=0.8$}. Left: entire angular coverage around a charge, showing regions of space affected by different states of charge motion due to a signal retardation effect (outer: rest; intermediate: acceleration; inner: uniform motion). Right: a selection of field lines within a limited angular range ahead of a charge, illustrating a portion of electric flux contained within a (quasi)conical opening. Shaded area indicates a closed integration surface, as described in the main text.} 
\label{fig2}
\end{figure*}

Let us parameterize a uniform motion of a point charge by a typical relativistic factor \mbox{$\boldsymbol{\beta}=\mathbf{v}/c$} (and its scalar form \mbox{$\beta=v/c$}), $\mathbf{v}$ being a charge velocity, with $c$ as a speed of light in vacuum.  A relativistically correct expression for its Coulomb (electric but no longer electrostatic) field reads:
\begin{equation}
\mathbf{E}(\mathbf{r},t)=\frac{q}{4\pi\epsilon_0}\frac{1-\beta^2}{(1-\beta^2\sin^2\Theta)^{3/2}}\frac{\hat{\mathbf{R}}}{R^2},
\label{uniform_field}
\end{equation}
where the time dependence enters through an evolving charge position \mbox{$\mathbf{r}_0(t)=\mathbf{r}_0(0)+ct\boldsymbol{\beta}$}. In that, $\Theta$ is an angle between a charge-relative position $\mathbf{R}$ and a charge velocity: \mbox{$\Theta=\sphericalangle(\mathbf{R},\boldsymbol{\beta})$}. This field is commonly represented by means of the `Lorentz contracted' lines field. An example of these is shown in the right panel from figure~\ref{fig1}, for a charge moving horizontally with \mbox{$\beta=0.9$}. An angular distribution of these lines is constructed in a following manner. For a total of $n$ field lines within a full angular range of $2\pi$, start with a set of uniformly distributed angles for a charge at rest: \mbox{$\theta_k=2\pi k/n$} for \mbox{$k=0,\ldots,n-1$}, wherein \mbox{$\theta_0=0$} corresponds to a direction of charge velocity. A set of `Lorentz contracted' angles $\theta'_k$ for the field lines of a uniformly moving charge is then obtained according to:
\begin{equation}
\tan\theta'_k=\gamma\tan(2\pi k/n);\quad k=0,\ldots,n-1,
\end{equation}
with $\gamma$ as a standard Lorentz factor: \mbox{$\gamma=(1-\beta^2)^{-1/2}$}. Evidently, an angular transformation for each particular field line is simply:
\begin{equation}
\tan\theta'=\gamma\tan\theta,
\label{master}
\end{equation}
with~$\theta'$ as a field line angle when charge is in motion and~$\theta$ as an angle of the `same' field line when charge is at rest. To our knowledge, there are two different arguments yielding the same transformation~(\ref{master}). One is by Feynman~\cite{feynman}. The other may be found in a paper by Tessman and Finnell~\cite{argument_2} or, for example, in a textbook by Purcell and Morin~\cite{purcell}.

\subsection{Feynman's argument}

Feynman's argument is a simple one~\cite{feynman}. It states that the field lines should be Lorentz contracted in a direction of charge motion, just as if they were real physical objects, subject to Lorentz transformations (but he makes it perfectly clear that they are not). Let us denote by~$\mathbf{f}$ a vectorial parametrization of a field line for a charge at rest, and by~$\mathbf{f}'$ a parametrization of the `same' field line for a charge in uniform motion. A straight field line segment~$\Delta \mathbf{f}$ between any two points $\mathbf{f}_2$ and $\mathbf{f}_1$ along a field line~$\mathbf{f}$ is \mbox{$\Delta \mathbf{f}=\mathbf{f}_2-\mathbf{f}_1$} (and analogously for~$\Delta \mathbf{f}'$). For simplicity, let us consider a two-dimensional space and a charge motion along the $x$-axis, so that \mbox{$\boldsymbol{\beta}=\beta\hat{\mathbf{x}}$}. In that case a (supposed) Lorentz contraction of a component parallel to a charge velocity, \mbox{$[\Delta \mathbf{f}']_x=[\Delta \mathbf{f}]_x/\gamma$}, and a trivial transformation of one perpendicular to it, \mbox{$[\Delta \mathbf{f}']_y=[\Delta \mathbf{f}]_y$}, lead to a ratio:
\begin{equation}
\frac{[\Delta \mathbf{f}']_y}{[\Delta \mathbf{f}']_x}=\gamma\frac{[\Delta \mathbf{f}]_y}{[\Delta \mathbf{f}]_x}.
\end{equation}
A field line angle~$\theta$ relative to a direction of charge motion is given by $\tan\theta=[\Delta \mathbf{f}]_y/[\Delta \mathbf{f}]_x$ (and analogously for~$\theta'$). Hence, this equation yields a transformation rule~(\ref{master}).

%angular deflection

\subsection{Flux argument}

An argument from~\cite{argument_2,purcell} relies on an electric field (i.e. an electric flux) satisfying the Gauss's law. We will therefore call it the \textit{flux argument}. It considers a point charge undergoing a \textit{rectilinear} acceleration from rest to a state of uniform motion, and examines the field lines at some moment after it has attained a constant velocity. Left panel from figure~\ref{fig2} shows an example of such field lines, for a final speed of $\beta=0.8$. Let us shortly describe a plot. A charge is assumed to be at rest, at a position~$\mathbf{r}_1$, until a moment~$t_1$. It accelerates until~\mbox{$t_2>t_1$}, attaining a final, constant velocity at~$\mathbf{r}_2$. Its electric field and the corresponding field lines are observed at some moment~\mbox{$t_3>t_2$}, when a chage is at~$\mathbf{r}_3$. For the purposes of flux argument the details of acceleration are irrelevant, save for the contraint that it needs to be rectilinear. For simplicity, figure~\ref{fig2} shows the field lines in a case of a relativistically uniform acceleration~\cite{rel_acc,rectilinear_acc}. At this point it is irrelevant how these field lines were calculated; the procedure will be explained in detail in Section~\ref{new_arg}. Due to a retardation effect (a finite time required for the information about a charge motion to reach faraway points), the `outer' field at distances greater than \mbox{$c(t_3-t_1)$} from the initial charge position~$\mathbf{r}_1$ is still electrostatic one. `Inner' field within a radius \mbox{$c(t_3-t_2)$} from a later position~$\mathbf{r}_2$ is a contracted Coulomb field due to a uniform motion. Intermediate field, between two dashed circles delineating an inner and outer region, is affected by a (past) charge acceleration and contains a component of electromagnetic radiation.

Flux argument relies only on an inner and outer field. The crux of the argument lies in considering a very specific closed integration surface passing through both the `inner' and `outer' region, while excluding the charge in order for a total integrated flux to be zero. An example of such surface is indicated by a shaded region from figure~\ref{fig2}. The surface in question corresponds to a two-dimensional boundary of a three-dimensional solid of revolution obtained by rotating a shaded area around the central (horizontal) axis. It comprises three relevant parts--- open surfaces $\Gamma_1$, $\Gamma_2$ and $\Gamma_3$---whose union forms a complete closed surface. Let~$\Gamma_1$ be a `cap' (not necessarily spherical nor necessarily perpendicular to an electric field) that is entirely contained within the `outer' region. It is bounded by all those field line segments (azimuthally distributed around the central axis) that close an angle~$\theta$ relative to a direction of a charge motion, so that a boundary of $\Gamma_1$ forms a circular outline intersecting all these $\theta$-segments. Similarly, let~$\Gamma_3$ be a cap entirely contained within the `inner' region, bounded by the field line segments defined by a sought angle $\theta'$. Most importantly, let~$\Gamma_2$ be a lateral quasi-conical surface \textit{spanned by the field lines}, connecting $\Gamma_1$ and $\Gamma_3$. By the definition of field lines, a field is parallel to~$\Gamma_2$ so that a contribution to the electric flux along this surface vanishes: \mbox{$\int_{\Gamma_2}\mathbf{E}\cdot\D\mathbf{A}=0$}. The Gauss's law then reduces to:
\begin{equation}
\oint\mathbf{E}\cdot\D\mathbf{A}=\int_{\Gamma_1}\mathbf{E}\cdot\D\mathbf{A}+\int_{\Gamma_3}\mathbf{E}\cdot\D\mathbf{A}=0.
\end{equation}
For simplicity we can now select the spherical caps perpendicular to a field. In that, $\Gamma_1$ is a cap from a sphere centred at~$\mathbf{r}_1$. A corresponding field to be integrated is the one from~(\ref{static_field}). Similarly, $\Gamma_3$ is a cap from a sphere centred at~$\mathbf{r}_3$; a field to be integrated now is a Coulomb field from~(\ref{uniform_field}). A careful arrangement of integrals leads to:
\begin{equation}
\int_0^\theta \sin\Theta \,\D\Theta=\int_0^{\theta'} \frac{1-\beta^2}{(1-\beta^2\sin^2\Theta)^{3/2}} \sin\Theta \,\D\Theta,
\label{gauss}
\end{equation}
wherein the term $\sin\Theta \,\D\Theta$ comes from using the spherical coordinates\footnote{
Though we may use two-dimensional coordinates for \textit{some} calculations involving electric fields, the familiar form of Gauss's law holds only in a three-dimensional space, hence a necessity of using the spherical, rather than polar coordinates.
} for a surface element~$\D\mathbf{A}$. Integration yields:
\begin{equation}
\cos\theta=\frac{\cos\theta'}{\sqrt{1-\beta^2\sin^2\theta'}},
\end{equation}
which, upon solving for~$\tan\theta'$, recovers~(\ref{master}).

\subsection{Discussion of arguments}

We are not strongly against these arguments. In fact, we quite appreciate Feynman's argument for its simplicity and elegance. However, this one has certain logical weaknesses. The flux argument---which we appreciate for its ingenious and deliciously shameless `exploitation' of the Gauss's law---features some limitations in what can be proven by it, which will become evident from the results of our own derivation.

Feynman's argument makes a connection between field lines from \textit{different observer frames}, as if they were real physical objects subject to Lorentz transformations. (He also makes it perfectly clear that they are not; that the argument is artificial and that obtained result is coincidental and miraculous.)

Flux argument provides a valid, though \textit{indirect} proof for a change in a field line inclination relative to a charge motion, and only for a charge undergoing a \textit{rectilinear} acceleration. The rectilinearity requirement is evident from a reliance of a calculation from~(\ref{gauss}) upon the \textit{axial symmetry} of the field lines piercing the integration caps~$\Gamma_1$ and $\Gamma_3$. Starting from an axially symmetric set of field lines `entering'~$\Gamma_3$, only in the case of a rectilinear acceleration can it be \textit{a priory} guaranteed that a set of field lines `exiting'~$\Gamma_1$ will also be axially symmetric. In case of a curvilinear acceleration it remains unclear how does a continual change in a velocity direction affect the final field line inclinations; whether there will be some nontrivial field line rotations (on top of the expected rotation induced by a varying velocity direction) or not. We will show---for an arbitrary \textit{planar} acceleration---that there will indeed be such additional rotations, making a general angular relation between the straight field line segments from the `inner' and `outer' region more complicated than~(\ref{master}). The flux argument still justifies representing a `Lorentz contracted' Coulomb field by the filed lines that are `Lorentz contracted' \textit{in a direction of a charge motion}, regardless of a past acceleration. The reason is that, once the uniform charge motion has been established, a Coulomb field (for a given charge velocity) does not depend on a history of charge acceleration. So the flux argument fully proves what it sets out to prove. But in \textit{general} case it fails to establish a connection between the particular field line segments from the `inner' and `outer' region---to which initial~$\theta$ does the final~$\theta'$ `belong'. Another aspect of the \textit{indirectness} of the flux argument consists in a \textit{magnitude} of the electric field having to satisfy the Gauss's law. Imagine an alternative vector field~$\mathbf{F}$ having at all points the same direction as an electric field, but arbitrarily modified values. For example: \mbox{$\mathbf{F}(\mathbf{r})=f(\mathbf{r})\mathbf{E}(\mathbf{r})$}, with \mbox{$f(\mathbf{r})>0$} as a strictly positive but otherwise arbitrary scalar field. Since the field lines depend only on a field \textit{direction} (see \ref{appendix_fl_eq}), this new field has \textit{exactly the same field lines} as an electric field, but it does not necessarily satisfy the Gauss's law. Therefore, exactly the same field lines can no longer be treated by the flux argument. Our proof will rely on the known geometry of the entire field line, and will be entirely independent of any reliance upon the field magnitude behavior.

Before proceeding to an improved argumentation, let us introduce a well known expression for an electric field of a point charge in an \textit{arbitrary} motion, so that the exposition of our argument can continue unimpeded. The expression reads~\cite{field1,field2,field3}:
\begin{equation}
\mathbf{E}(\mathbf{r},t)=\frac{q}{4\pi\epsilon_0}\left[\frac{\hat{\mathbf{R}}-\boldsymbol{\beta}}{\gamma^2K^3R^2}+\frac{\hat{\mathbf{R}}\times((\hat{\mathbf{R}}-\boldsymbol{\beta})\times\mathbf{a})}{c^2K^3R}\right]_\tau .
\label{E_general}
\end{equation}
Most of the terms have been defined by the point of introducing~(\ref{uniform_field}). Newly appearing terms are a charge acceleration \mbox{$\mathbf{a}=\D\mathbf{v}/\D t$} and \mbox{$K=1-\hat{\mathbf{R}}\cdot\boldsymbol{\beta}$}. The content of the brackets~$[\cdot]_\tau$ is to be evaluated at the retarded time~$\tau$ such that \mbox{$\tau+R(\tau)/c=t$}, due to a finite speed of the information transfer between a charge position~$\mathbf{r}_0$ and a field observation point~$\mathbf{r}$. In a previous work~\cite{point_charge} we have demonstrated that~(\ref{E_general}) indeed reduces to~(\ref{uniform_field}) in case of a uniform charge motion (for constant~$\boldsymbol{\beta}$ and \mbox{$\mathbf{a}=\mathbf{0}$}). There is also a Feynman's formula~\cite{field4,field5} equivalent to~(\ref{E_general}):
\begin{equation}
\mathbf{E}(\mathbf{r},t)=\frac{q}{4\pi\epsilon_0}\left(\bigg[\frac{\hat{\mathbf{R}}}{R^2}\bigg]_\tau +\frac{[R]_\tau}{c}\frac{\D}{\D t}\bigg[\frac{\hat{\mathbf{R}}}{R^2}\bigg]_\tau+\frac{1}{c^2}\frac{\D^2[\hat{\mathbf{R}}]_\tau}{\D t^2} \right),
\end{equation}
but we do not pursue it further in this work.

\section{Improved proof}
\label{new_arg}

We improve upon the idea of the flux argument by providing a \textit{direct} proof for a `field line contraction'\ from~(\ref{master}). Just like a proof of the flux argument, our proof relies on a continuity of field lines between two states of uniform motion. However, instead of the Gauss's law we use a little known, half a century old result by Tsien~\cite{general_acc}: an exact analytical parametrization of the electric field lines for a point charge in \textit{arbitrary planar} motion. The proof provides a generalization of the `angular contraction' from~(\ref{master}), showing that the straight field line segments may undergo additional rotation in case of the curvilinear charge acceleration. Another aspect of the directness of our proof is the fact that it relies on a continuity of \textit{each particular field line}. The flux argument---by considering an electric flux through an angular opening bounded by a family of field lines---relies on a continuity of the entire set of field lines.

Let us restate the basics of the argument. Consider a charge that accelerates from rest to a state of uniform motion. Draw the electric field lines at some moment during this uniform motion. Just as in figure~\ref{fig2}, there will be an outer region of space where the field lines are still affected by a past rest state, and an inner region where they are affected by a present uniform motion (separated by a region with electromagnetic radiation due to a charge acceleration). Within an outer region select the field line segments isotropically, so that within this rest-affected-region they have a uniform angular separation. There can hardly be any argument against this selection for the field lines of a charge at rest, as their isotropy represents an isotropy of an electrostatic field. Now, whatever the form of acceleration from rest may be, trace the field lines toward the charge. In other words, extend them toward an inner region where a field is affected by uniform motion. Do so by solving a field line equation (see \ref{appendix_fl_eq}):
\begin{equation}
\frac{\D \mathbf{f}}{\D\ell}=\pm\hat{\mathbf{E}}(\mathbf{f})
\label{fl_equation}
\end{equation}
for an electric field from~(\ref{E_general}), produced by a selected form of acceleration. A sign in front of a unit field~$\hat{\mathbf{E}}$ should be selected on a basis of a charge~$q$, so that the field lines lead toward the charge, rather than away from it.

Before proceeding to a general proof, we note here a more specific result by Ruhlandt et al.~\cite{rectilinear_acc}.  They have shown, by applying their findings for an arbitrary \textit{rectilinear} motion, that if a field line of a charge initially at rest points along a direction $\hat{\mathbf{R}}$, after arbitrary rectilinear acceleration to a uniform motion with a velocity~$\boldsymbol{\beta}$ it will point along:
\begin{equation}
\hat{\mathbf{R}}'=\frac{(\gamma^{-1}-1)(\hat{\mathbf{R}}\cdot\hat{\boldsymbol{\beta}})\hat{\boldsymbol{\beta}}+\hat{\mathbf{R}}}{\sqrt{1-(\hat{\mathbf{R}}\cdot\boldsymbol{\beta})^2}}.
\label{rectilinear}
\end{equation}
This is equation~(16) from~\cite{rectilinear_acc}, adjusted for our notation. Multiplying (as a dot product) this equation by~$\hat{\boldsymbol{\beta}}$, while using \mbox{$\hat{\mathbf{R}}\cdot\hat{\boldsymbol{\beta}}=\cos\theta$} and \mbox{$\hat{\mathbf{R}}'\cdot\hat{\boldsymbol{\beta}}=\cos\theta'$}, we obtain:
\begin{equation}
\cos\theta'=\frac{\cos\theta}{\sqrt{\gamma^2\sin^2\theta+\cos^2\theta}}=\frac{\cos\theta}{\gamma\sqrt{1-\beta^2\cos^2\theta}}.
\label{coss}
\end{equation}
Using elementary trigonometry this is easily shown to be equivalent to~(\ref{master}), immediately providing a direct geometric proof circumventing the Gauss's law.

\subsection{On with a proof}

Using half a century old result by Tsien~\cite{general_acc}, we will now prove that the electric field lines of a point charge initially at rest are indeed  `Lorentz contracted' in a direction of charge motion, after an \textit{arbitrary} planar acceleration to a state of uniform motion. However, for a general curvilinear acceleration---as opposed to a general rectilinear acceleration, yielding~(\ref{rectilinear})---there will be a nontrivial rotation of relevant field line segments, relative to a direction of charge motion.

\begin{figure}[t!]
\centering
\includegraphics[width=0.45\textwidth,keepaspectratio]{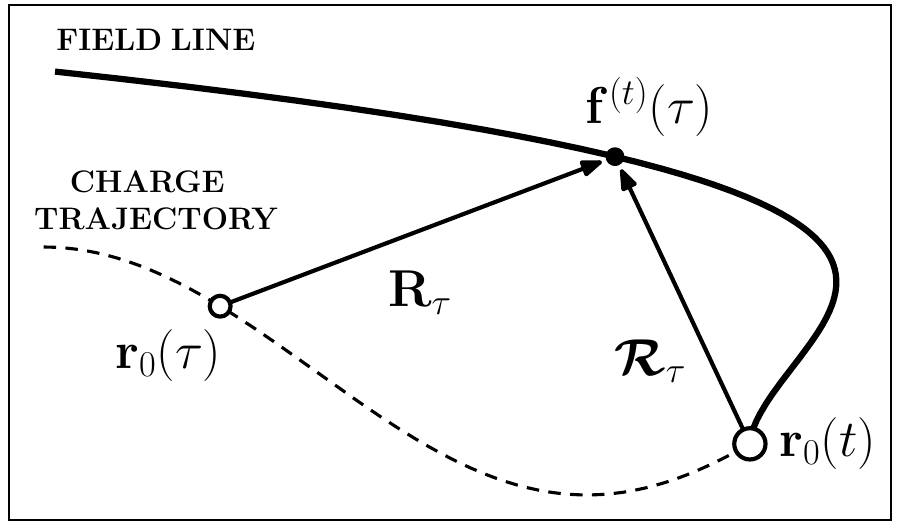}
\caption{Meaning of geometric coordinates from~(\ref{R_ret}) and~(\ref{R_pres}). Retarded time is~$\tau$; present time is~$t$.}
\label{fig3}
\end{figure}

Tsien~\cite{general_acc} derives an analytical field line parametrization by employing a special set of curvilinear coordinates, while using a retarded time~$\tau$ in order to parameterize the entire field line \textit{at present time~$t$}. For this reason we denote a field line itself by $\mathbf{f}^{(t)}$ and a particular point on it by $\mathbf{f}^{(t)}(\tau)$, the point itself being parameterized by~$\tau$. Using a simplified notation $\mathbf{R}_\tau=\mathbf{R}(\tau)$ for representing a time dependence for quantities that will figure prominently in the following expressions, we introduce the following terms:
\begin{align}
&\mathbf{R}_\tau=\mathbf{f}^{(t)}(\tau)-\mathbf{r}_0(\tau),
\label{R_ret}\\
&\boldsymbol{\mathcal{R}}_\tau=\mathbf{f}^{(t)}(\tau)-\mathbf{r}_0(t).
\label{R_pres}
\end{align}
The first one is a position~$\mathbf{R}_\tau$ of a specific field line point relative to a \textit{retarded} charge position~$\mathbf{r}_0(\tau)$. The second one is a position~$\boldsymbol{\mathcal{R}}_\tau$ of the same point relative to a \textit{present} charge position~$\mathbf{r}_0(t)$. Figure~\ref{fig3} illustrates a geometric meaning of these coordinates, analogously to figure~1 from~\cite{general_acc}. Both~$\mathbf{R}_\tau$ and~$\boldsymbol{\mathcal{R}}_\tau$ also depend on a present time~$t$, but we suppress this dependence for a simplicity of notation.

Equation~(3) from~\cite{general_acc} parameterizes field line coordinates using a retarded position from~(\ref{R_ret}):
\begin{align}
&[\mathbf{f}^{(t)}(\tau)]_x=x_0(\tau)+R_\tau\cos(\varphi_\tau+\alpha_\tau),
\label{fl_x}\\
&[\mathbf{f}^{(t)}(\tau)]_y=y_0(\tau)+R_\tau\sin(\varphi_\tau+\alpha_\tau),
\label{fl_y}
\end{align}
with $x_0$ and $y_0$ as the components of a charge position~$\mathbf{r}_0$. In that, we freely alter a notation from~\cite{general_acc}, adapting it to this work. Tsien manages to analytically parameterize the field lines by separating a total angle 
\mbox{$\varphi_\tau+\alpha_\tau$} between $\mathbf{R}_\tau$ and the $x$-axis into: (1)~an angular deflection~$\varphi_\tau$ between $\boldsymbol{\beta}_\tau$ and $x$-axis, and (2)~an angular deflection~$\alpha_\tau$ between~$\boldsymbol{\beta}_\tau$ and~$\mathbf{R}_\tau$. In summary:
\begin{align}
&\varphi_\tau+\alpha_\tau=\sphericalangle(\mathbf{R}_\tau,\hat{\mathbf{x}}),
\label{varphi_alpha}\\
&\varphi_\tau=\sphericalangle(\boldsymbol{\beta}_\tau,\hat{\mathbf{x}}),
\label{varphi}\\
&\alpha_\tau=\sphericalangle(\mathbf{R}_\tau,\boldsymbol{\beta}_\tau).
\label{alpha}
\end{align}
In that, $R_\tau$, $\varphi_\tau$ and $\alpha_\tau$ are treated as a special set of curvilinear coordinates. By definition, a norm of a retarded quantity~$\mathbf{R}_\tau$---the same one as in~(\ref{E_general})---is:
\begin{equation}
R_\tau=(t-\tau)c.
\label{r_tau}
\end{equation}
At any point, $\varphi_\tau$ is easily determined from a charge motion, as per definition from~(\ref{varphi}): \mbox{$\cos\varphi_\tau=\hat{\boldsymbol{\beta}}_\tau \cdot\hat{\mathbf{x}}$}. A nontrivial component is~$\alpha_\tau$, whose analytical identification for a general charge motion is a pinnacle of~\cite{general_acc}:
\begin{equation}
\tan\case{1}{2}\alpha_\tau=\sqrt{\frac{1-\beta_\tau}{1+\beta_\tau}} \tan\left[\frac{\vartheta}{2}+\frac{1}{2}\int_\tau^t \gamma(\mathcal{T}) \omega(\mathcal{T}) \D \mathcal{T}\right].
\label{tan_alpha}
\end{equation}
This is equation~(13) from~\cite{general_acc}, with a notation slightly adjusted to this work. Angular parameter~$\vartheta$ is an integration constant defining a particular field line. As usual, $\gamma$ is the Lorentz factor: \mbox{$\gamma(\mathcal{T})=(1-\beta_\mathcal{T}^2)^{-1/2}$}, with $\omega$ as a rate of change of an angular coordinate $\varphi$, that may be expressed as:
\begin{equation}
\omega(\mathcal{T})=\frac{\D \varphi_\mathcal{T}}{\D\mathcal{T}}=\frac{(\boldsymbol{\beta}_\mathcal{T}\times \mathbf{a}_\mathcal{T}/c )\cdot\hat{\mathbf{z}}}{\beta_\mathcal{T}^2},
\label{omega}
\end{equation}
with $\mathbf{a}_\mathcal{T}$ as a charge acceleration: \mbox{$\mathbf{a}_\mathcal{T}=\D \mathbf{v}_\mathcal{T}/\D\mathcal{T}$}.

We now exclusively focus on our case of interest. We examine an electric field of a point charge that is initially at rest, then undergoes an arbitrary planar acceleration within a time interval \mbox{$0\le \tau\le T$}, ending in a uniform motion. In summary: %with \mbox{$\boldsymbol{\beta}_{t\ge T}=\frac{1}{c}\int_0^T \mathbf{a}_\mathcal{T} \D \mathcal{T}$}
\begin{equation*}
\begin{array}{lll}
\tau<0: & \mbox{charge at rest} & \big[ \boldsymbol{\beta}_\tau=\mathbf{0}\big],\\[\smallskipamount]
0\le \tau \le T: & \mbox{arbitrary acceleration} & \big[\boldsymbol{\beta}_\tau=\frac{1}{c}\int_0^\tau \mathbf{a}_\mathcal{T} \D \mathcal{T}\big],\\[\smallskipamount]
\tau>T: & \mbox{uniform motion} & \big[\boldsymbol{\beta}_\tau=\frac{1}{c}\int_0^T \mathbf{a}_\mathcal{T} \D \mathcal{T}\big].
\end{array}
\end{equation*}
In that, we observe some moment \mbox{$t>T$} when a charge has attained a constant velocity~$\boldsymbol{\beta}_T$. For an illustration of such case, a reader may refer to figure~\ref{fig2} or a later figure~\ref{fig4}. In accordance with our argument, select a starting point for tracing a particular field line within an \textit{outer} region that is still affected by the initial electrostatic field of a charge at rest. We wish to deduce a relation between the \textit{external} field line angle~$\theta_\mathrm{ext}$ from this outer region (field line segment parameterized by \mbox{$\tau<0$}) and the \textit{internal} field line angle~$\theta_\mathrm{int}$ from the inner region affected by a Coulomb field of a uniformly moving charge (field line segment parameterized by \mbox{$\tau>T$}). We wish that these angles bear a correspondence to~$\theta$ and~$\theta'$ from~(\ref{master}) and~(\ref{coss}): \mbox{$\theta_\mathrm{ext}\leftrightarrow \theta$} and \mbox{$\theta_\mathrm{int}\leftrightarrow \theta'$}. Therefore, they must be defined relative to a charge velocity (as opposed to some fixed direction), since the sought angle transformation comes from the asymmetry of Lorentz transformations, relative to a direction of motion. It is now imperative to precisely specify these angles. Since the starting~$\theta_\mathrm{ext}$ should retain its definition independently of the subsequent charge motion, it should be defined relative to a \textit{retarded} charge motion prior to acceleration. If it were defined relative to a \textit{present} motion, it would be affected by an arbitrary direction of a final velocity. On the other hand, $\theta_\mathrm{int}$ should be defined relative to a present motion, i.e. relative to~$\boldsymbol{\beta}_t=\boldsymbol{\beta}_T$. Since it is assumed that \mbox{$\boldsymbol{\beta}_T\neq\mathbf{0}$}, i.e. a direction of $\boldsymbol{\beta}_T$ is well defined, there are no further issues regarding $\theta_\mathrm{int}$. However, $\theta_\mathrm{ext}$ comes from the initial rest state, when \mbox{$\boldsymbol{\beta}_\tau=\mathbf{0}$}, so that a direction of~$\boldsymbol{\beta}_\tau$ is not well defined. For this reason we select a direction of an initial velocity increment~$\D\boldsymbol{\beta}_0$, that obviously corresponds to a direction of initial acceleration~$\mathbf{a}_0$.

It just remains to identify which vectorial quantity carries a sought field line direction. Since \textit{straight field line segments}---for which a field line direction can be meaningfully defined---`travel' with the charge, their direction is defined relative to a \textit{last charge position} from which these straight segments originate or have originated. In our setup there are two such families of straight segments. One is composed of inner segments actively originating from a uniformly moving charge. Therefore, their direction is defined relative to a \textit{present} charge position~$\mathbf{r}_0(t)$ appearing in~(\ref{R_pres}). Hence:
\begin{equation}
\theta_\mathrm{int}=\sphericalangle(\boldsymbol{\mathcal{R}}_\tau,\boldsymbol{\beta}_T) \quad\mbox{for}\quad T<\tau<t.
\label{theta_int}
\end{equation}
The other family is composed of outer segments, last affected at \mbox{$\tau=0$} by a charge at rest at a \textit{retarded} position~$\mathbf{r}_0(0)$. Since \mbox{$\mathbf{r}_0(\tau)=\mathbf{r}_0(0)$} for \mbox{$\tau\le0$}, due to a continual rest, we may use~(\ref{R_ret}) to define:
\begin{equation}
\theta_\mathrm{ext}=\sphericalangle\big(\mathbf{R}_\tau,\mathbf{a}_0\big) \quad\mbox{for}\quad \tau<0.
\label{theta_ext}
\end{equation}

Now the question is: how is an integration constant~$\vartheta$ from~(\ref{tan_alpha}) related to~$\theta_\mathrm{ext}$ from an outer, electrostatic region? We find this simply by applying~(\ref{tan_alpha})---at a present moment \mbox{$t>T$}--- to some retarded moment $\tau<0$. Due to \mbox{$\beta_\tau=0$} an entire square root reduces to unity. For a \textit{charge at rest} the angle~$\alpha_\tau$ precisely corresponds to a field line angle~$\theta_\mathrm{ext}$, when a direction of $\boldsymbol{\beta}_\tau$ from~(\ref{alpha}) is exchanged for a direction of initial acceleration~$\mathbf{a}_0$. The reason for this is the same as in~(\ref{theta_ext}): for a charge at rest a direction of~$\boldsymbol{\beta}_\tau$ is undefined. We remind a reader that a direction of~$\mathbf{a}_0$ only serves as a stand-in for a direction of the initial velocity increment~$\D\boldsymbol{\beta}_0$, so that no fundamental change in a definition of~$\alpha_\tau$ has occurred. The selection of~$\mathbf{a}_0$ as a substitute for~$\boldsymbol{\beta}_\tau$ is justified by the \textit{continuity of reference direction} for gauging~$\mathbf{R}_\tau$, i.e. by the continuity of a transition between~$\hat{\mathbf{a}}_0$ to~$\hat{\boldsymbol{\beta}}_\tau$. Finally, for \mbox{$t>T$} and \mbox{$\tau<0$} (precisely the case we are observing), an integral from~(\ref{tan_alpha}) is contributed by an \textit{entire} accelerated motion, since the acceleration is confined within a time interval from~0 to~$T$. Under these circumstances an integral simply represents a constant contribution~$\Delta \theta$:
\begin{equation}
\Delta\theta=\int_0^T \gamma(\mathcal{T}) \omega(\mathcal{T}) \D \mathcal{T}.
\label{delta_theta}
\end{equation}
Thus, for \mbox{$t>T$} and \mbox{$\tau<0$} the entirety of~(\ref{tan_alpha}) reduces to:
\begin{equation}
\tan \case{1}{2}\theta_\mathrm{ext}=\tan \case{1}{2}(\vartheta+\Delta\theta),
\end{equation}
which yields (after some obvious continuity arguments applied to the terms under the tangent functions):
\begin{equation}
\vartheta=\theta_\mathrm{ext}-\Delta\theta.
\label{vartheta}
\end{equation}
This is a sought connection between~$\vartheta$ and $\theta_\mathrm{ext}$, which can be taken back to~(\ref{tan_alpha}).

We now apply~(\ref{tan_alpha}) to some retarded moment \mbox{$\tau>T$}, after the acceleration has ended and a final velocity~$\boldsymbol{\beta}_T$ has been attained. The corresponding field line segments are the straight line segments from an inner region, centered at a \textit{present} charge position~$\mathbf{r}_0(t)$. Since the acceleration has ended, an angular rate of change $\omega(\mathcal{T})$ from~(\ref{omega}) vanishes for \mbox{$T<\tau\le\mathcal{T}\le t$}. So does an integral from~(\ref{tan_alpha}), yielding for \mbox{$\tau>T$}:
\begin{equation}
\tan\case{1}{2}\alpha_\tau=\sqrt{\frac{1-\beta_T}{1+\beta_T}} \tan \case{1}{2}\vartheta.
\label{tan_alpha_new}
\end{equation}
For a charge in motion the angle~$\alpha_\tau$ does not correspond to a field line angle from~(\ref{theta_int}), since $\theta_\mathrm{int}$ is defined relative to a present charge position~$\mathbf{r}_0(t)$, as opposed to a retarded position~$\mathbf{r}_0(\tau)$ used for~$\alpha_\tau$. In case of a uniform motion with a constant~$\boldsymbol{\beta}_T$, their relation is trivial:
\begin{equation}
\mathbf{r}_0(t)-\mathbf{r}_0(\tau)=(t-\tau)c\boldsymbol{\beta}_T.
\label{dr_t_tau}
\end{equation}
We now need to establish a connection between~$\theta_\mathrm{int}$ and~$\alpha_\tau$. With some elementary trigonometry~(\ref{tan_alpha_new}) is easily translated into:
\begin{align}
&\sin\alpha_\tau=\dfrac{\sin\vartheta}{\gamma_T(1+\beta_T\cos\vartheta)},
\label{sin_alpha}\\
&\cos\alpha_\tau=\dfrac{\beta_T+\cos\vartheta}{1+\beta_T\cos\vartheta}.
\label{cos_alpha}
\end{align}
Due to a requirement for~$\boldsymbol{\mathcal{R}}_\tau$ from~(\ref{theta_int}), we eliminate $\mathbf{f}^{(t)}(\tau)$ from~(\ref{R_ret}) and~(\ref{R_pres}) so as to obtain:
\begin{equation}
\boldsymbol{\mathcal{R}}_\tau=\mathbf{R}_\tau - [\mathbf{r}_0(t)-\mathbf{r}_0(\tau)]=\mathbf{R}_\tau-(t-\tau)c\boldsymbol{\beta}_T,
\label{RRrr}
\end{equation}
where the last equality was obtained by using~(\ref{dr_t_tau}). Writing out the components of~$\boldsymbol{\mathcal{R}}_\tau$---using the angular deflections for $\mathbf{R}_\tau$ and $\boldsymbol{\beta}_T$ from~(\ref{varphi_alpha}) and~(\ref{varphi})---gives:
\begin{align}
&[\boldsymbol{\mathcal{R}}_\tau]_x=R_\tau\cos[\varphi_T+\alpha_\tau]-(t-\tau)\beta c\,\cos\varphi_T,\\
&[\boldsymbol{\mathcal{R}}_\tau]_y=R_\tau\sin[\varphi_T+\alpha_\tau]-(t-\tau)\beta c\,\sin\varphi_T.
\end{align}
Expanding the trigonometric terms and using~(\ref{r_tau}) yields:
\begin{align}
&[\boldsymbol{\mathcal{R}}_\tau]_x=(t-\tau)c\left[\big(\cos\alpha_\tau-\beta_T\big)\cos\varphi_T - \sin\alpha_\tau \sin\varphi_T\right],\\
&[\boldsymbol{\mathcal{R}}_\tau]_y=(t-\tau)c\left[\big(\cos\alpha_\tau-\beta_T\big)\sin\varphi_T + \sin\alpha_\tau \cos\varphi_T\right].
\end{align}
With~(\ref{sin_alpha}) and~(\ref{cos_alpha}) we finally arrive at:
\begin{align}
&[\boldsymbol{\mathcal{R}}_\tau]_x=\dfrac{\cos\varphi_T\cos\vartheta-\gamma_T\sin\varphi_T\sin\vartheta}{\gamma_T^2(1+\beta_T\cos\vartheta)}
\label{R_tau_x},\\
&[\boldsymbol{\mathcal{R}}_\tau]_y=\dfrac{\sin\varphi_T\cos\vartheta+\gamma_T\cos\varphi_T\sin\vartheta}{\gamma_T^2(1+\beta_T\cos\vartheta)}
\label{R_tau_y}.
\end{align}
Using \mbox{$\hat{\boldsymbol{\beta}}_T=\cos\varphi_T\, \hat{\mathbf{x}}+\sin\varphi_T\, \hat{\mathbf{y}}$}, we obtain the sought relation for~$\theta_\mathrm{int}$ according to~(\ref{theta_int}):
\begin{equation}
\cos\theta_\mathrm{int}=\hat{\boldsymbol{\mathcal{R}}}_\tau\cdot\hat{\boldsymbol{\beta}}_T=\frac{\cos\vartheta}{\gamma_T\sqrt{1-\beta_T^2\cos^2\vartheta}}.
\end{equation}
In analogy with the equivalence between~(\ref{master}) and~(\ref{coss}), we immediately conclude that this result is equivalent\footnote{
We may also arrive at the same conclusion by recognizing that the components from~(\ref{R_tau_x}) and~(\ref{R_tau_y}) have a form of a vector rotated by an angle~$\varphi_T$:
\begin{equation*}
\boldsymbol{\mathcal{R}}_\tau=\frac{1}{\gamma_T^2(1+\beta_T\cos\vartheta)}
\left[\begin{array}{cc}
\cos\varphi_T & -\sin\varphi_T\\
\sin\varphi_T & \cos\varphi_T
\end{array}\right]
\left[\begin{array}{c}
\cos\vartheta\\
\gamma_T\sin\vartheta
\end{array}\right].
\label{rotated}
\end{equation*}
Since~$\varphi_T$ is precisely the angle of $\boldsymbol{\beta}_T$, the components of a vector being rotated (\mbox{$\cos\vartheta\,\hat{\mathbf{x}}+\gamma_T\sin\vartheta\,\hat{\mathbf{y}}$}) quickly yield the sought angle~$\theta_\mathrm{int}$ relative to~$\boldsymbol{\beta}_T$. In order to formalize the statement, let~$\hat{\boldsymbol{\beta}}_T$ and~\mbox{$\hat{\boldsymbol{\lambda}}_T=\hat{\mathbf{z}}\times\hat{\boldsymbol{\beta}}_T$} be orthogonal basis vectors.  Then we may write: %Based on~(\ref{rotated}) we may write:
\begin{equation*}
\boldsymbol{\mathcal{R}}_\tau=\frac{\cos\vartheta \,\hat{\boldsymbol{\beta}}_T + \gamma_T\sin\vartheta \,\hat{\boldsymbol{\lambda}}_T}{\gamma_T^2(1+\beta_T\cos\vartheta)}.
\end{equation*}
The tangent of~$\theta_\mathrm{int}$ is given by the ratio of components in this basis. 
} to \mbox{$\tan\theta_\mathrm{int}=\gamma_T \tan\vartheta$}. Using~(\ref{vartheta}) we have:
\begin{equation}
\tan\theta_\mathrm{int}=\gamma_T \tan(\theta_\mathrm{ext}-\Delta\theta).
\label{proof}
\end{equation}
This is a sought relation between the angles of the straight field line segments produced by a charge at rest~($\theta_\mathrm{ext}$) and the same charge accelerated to a uniform motion~($\theta_\mathrm{int}$), concluding our proof.

\pagebreak

\subsection{Discussion of the result}

The result~(\ref{proof}) clearly shows that, in general, there is a nontrivial rotation of the straight field line segments, by an angle~$\Delta\theta$ \textit{relative to the final velocity}~$\boldsymbol{\beta}_T$. A trivial rotation \textit{with the velocity}, i.e. due to a rotation of the velocity vector itself by the angle \mbox{$\varphi_T-\varphi_0$}, is already accounted by the definition of~$\theta_\mathrm{int}$ and $\theta_\mathrm{ext}$. This means that \textit{prior to applying the contraction} by factor~$\gamma_T$ from~(\ref{proof}), the initial field line segments are rotated by a total angle \mbox{$\Theta_0=\varphi_T-\varphi_0-\Delta\theta$}, which is the same for \textit{all} field lines. After the contraction has been applied, the total rotation of a straight segment produced by a uniform motion is dependent on a particular field line and amounts to \mbox{$\Theta=(\varphi_T+\theta_\mathrm{int})-(\varphi_0+\theta_\mathrm{ext})$}. Neither the additional rotation by~$\Delta\theta$ nor the total rotation of the initial field line segments by~$\Theta_0$ is present in case of a rectilinear acceleration, as confirmed by a specific result from~(\ref{rectilinear}) and~(\ref{coss}). Within a general framework this is direct consequence of \mbox{$\omega=0$} from~(\ref{omega}).

It is well known that back-to-back field lines of a charge at rest---i.e. those lying on the same line and connecting to a charge with a relative angular deflection of $\pi$---remain back-to-back after a pure contraction by means of~(\ref{master}). Since, in general case, the `preliminary' rotation of the initial field line segments by~$\Theta_0$ is the same for all field lines, this rotation preserves their back-to-back property prior to contraction. Therefore, even with a nontrivial~$\Delta\theta$, initially back-to-back field line segments will still remain back-to-back after contraction. In other words, if the charge acceleration translates the initial field line angle~$\theta_\mathrm{ext}$ into~$\theta_\mathrm{int}$, then it translates the initial angle \mbox{$\theta'_\mathrm{ext}=\theta_\mathrm{ext}+\pi$} into \mbox{$\theta'_\mathrm{int}=\theta_\mathrm{int}+\pi$}. Stated in a functional dependence notation:
\begin{equation}
\theta_\mathrm{int}(\theta_\mathrm{ext}+\pi)=\theta_\mathrm{int}(\theta_\mathrm{ext})+\pi.
\label{back_to_back}
\end{equation}
This conclusion is supported by applying a trigonometric identity \mbox{$\tan(\theta+\pi)=\tan\theta$} to~(\ref{proof}).

In order to demonstrate this nontrivial rotation effect, we consider a specific example of a uniform \textit{tangential} acceleration along the circle. At $\tau=0$ a charge initially at rest undergoes a constant tangential acceleration until attaining a speed~$\beta_T$ at $\tau=T$, while circumscribing a total angle~$\varphi_T$ along the circle. The total acceleration is neither tangential nor centripetal, hence no \textit{specific} result from~\cite{general_acc} applies and we must make use of a general procedure from~(\ref{tan_alpha}). In that, we wish to control a selection of~$\beta_T$ and~$\varphi_T$. For a given~$T$ this uniquely determines a radius of a circular trajectory as  \mbox{$\rho=Tc\beta_T/2\varphi_T$}. During an acceleration interval \mbox{$0\le\tau\le T$} the velocity vector \mbox{$\boldsymbol{\beta}_\tau=\beta_\tau(\cos\varphi_\tau\, \hat{\mathbf{x}}+\sin\varphi_\tau\, \hat{\mathbf{y}})$} is parameterized by \mbox{$\beta_\tau=(\tau/T)\beta_T$} and \mbox{$\varphi_\tau=(\tau/T)^2\varphi_T$}.

\begin{figure}[t!]
\centering
\includegraphics[width=0.35\textwidth,keepaspectratio]{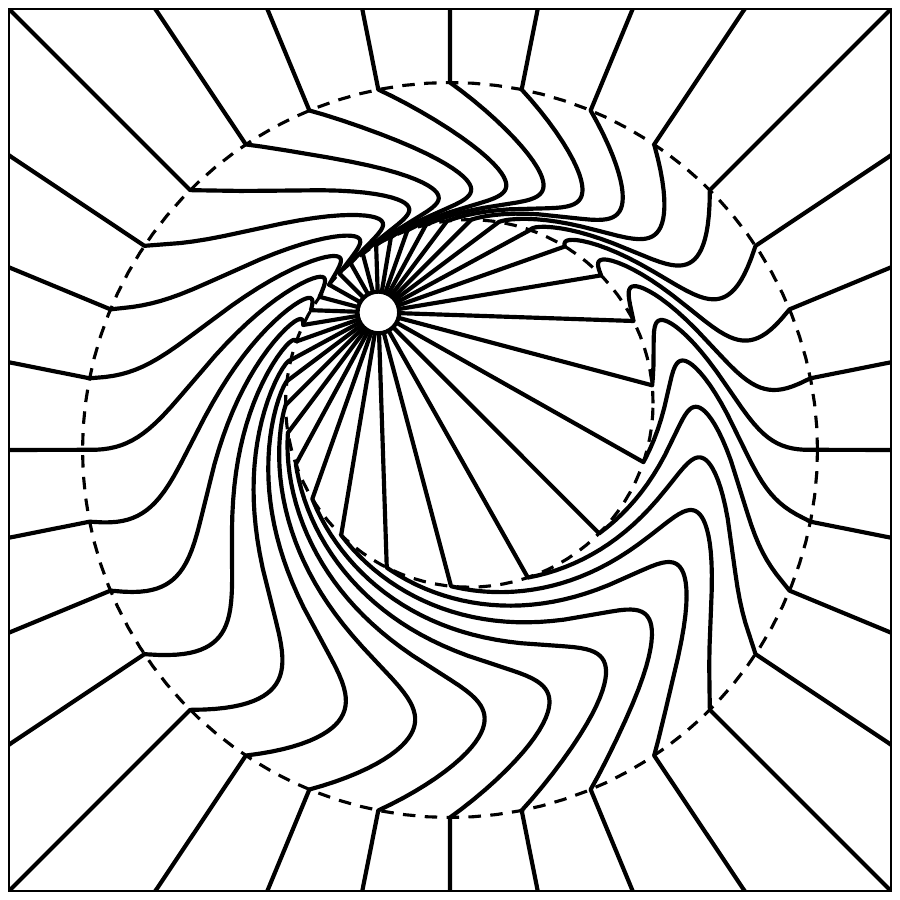}
%\vspace*{-1mm}
\caption{Field lines of a charge initially at rest (segments from an external region), that undergoes a uniform tangential acceleration along the circle, in a counter-clockwise direction (producing the intermediate curvilinear segments), culminating in a uniform motion with \mbox{$\beta_T=0.7$} and and a total angular deflection \mbox{$\varphi_T=3\pi/4$} (segments from an internal region).}
%\vspace*{-1mm}
\label{fig4}
\end{figure}

Figure~\ref{fig4} shows the field lines at some moment \mbox{$t>T$} when the charge drifts uniformly with~{$\beta_T=0.7$}, along a straight line deflected by \mbox{$\varphi_T=3\pi/4$} (roughly towards the upper left corner, due to a plot being centered at the initial charge position---a convergence point of the external field line segments). An angular deflection~$\Delta\theta$ from~(\ref{delta_theta}) may be analytically calculated, yielding \mbox{$\Delta\theta=2(1-\gamma_T^{-1})\beta_T^{-2}\varphi_T$}. In our case it amounts to approximately \mbox{$\Delta\theta\approx7\pi/8$}. A rotation effect is best seen by tracing a pair of externally horizontal lines toward the charge. Their angles within the inner region, calculated from~(\ref{proof}), are \mbox{$\theta_\mathrm{int}(\theta_\mathrm{ext}=0)=-0.833\pi$} and \mbox{$\theta_\mathrm{int}(\theta_\mathrm{ext}=\pi)=0.167\pi$}, in agreement with~(\ref{back_to_back}). It can be clearly seen that the initially forward line (\mbox{$\theta_\mathrm{ext}=0$}) is backward directed (relative to a present charge velocity) within the inner region, while the initially backward line (\mbox{$\theta_\mathrm{ext}=\pi$}) becomes forward directed.

\section{Conclusions}
\label{conclusions}

We have addressed a common practice of depicting a Coulomb field of a point charge in uniform motion by means of `Lorentz contracted' fields lines. Since there is no physical reality to the field lines, using them for representing any physical property (of the vector field) should be well argued and logically justified. We have presented two existing arguments: one based on the Lorentz contraction of geometric coordinates between different observer frames (Feynman's argument), the other one based on the Gauss's law for an electric field in vacuum (the flux argument). Based on a continuity of individual field lines we have provided an improved proof for the `Lorentz contraction' of the fields lines, yielding a generalization of the past result. Though a new proof is mathematically more difficult to demonstrate than a proof of the conceptually similar flux argument, the required calculations are still well within a grasp of any (under)graduate student. We do recognize that the greater effort required to complete these calculations may not make them a typical `weapon of choice'. For this reason we do not propose to abandon the previous argumentations altogether. Our suggestion is to present the students with multiple arguments, together with a clarification of their strengths, weaknesses and/or limitations. After clearly articulating that they all lead to the same (basic) result, a computational simplicity of past arguments may still be used for a convincing demonstration of the field lines contraction. But the issue of logical foundation and the applicability of these calculations should be clearly conveyed and understood.

\section*{Acknowledgments}

The work of I.S. was supported by the Croatian Science Foundation Project No. IP-2020-02-9614.

\appendix

\section{Field line equation}
\label{appendix_fl_eq}

Though a general field line equation~(\ref{fl_equation}) seems to be well known, we have not been able to find a well documented rationale behind it. This is also partly reflected in a plethora of its equivalent forms that may be found throughout the literature \cite{electric_1,magnetic_1,magnetic_3,magnetic_4,rectilinear_acc,general_acc,dipole,plotting}, without any centralized reference to them. Since a general solution to this equation---encompassed by~(\ref{fl_x}) and~(\ref{fl_y})---is of central importance to our proof, we provide here a \textit{well motivated and logically justified} introduction of a general field line equation, rather than posing it `out of nowhere' in its final form.

Let us treat a particular field line of a vector field $\mathbf{E}(\mathbf{r})$ (electric or otherwise) as a parametric curve $\mathbf{f}(\ell)$, parameterized by some (unspecified) parameter~$\ell$. The only requirement that we impose upon it is that~$\mathbf{f}$ be tangential to a vector field. Thus, we may write:
\begin{equation}
\frac{\D \mathbf{f}(\ell)}{\D\ell}=\lambda\mathbf{E}[\mathbf{f}(\ell)],
\label{fl_start}
\end{equation}
with~$\lambda$ as some, as yet unspecified factor. From this point on we omit a dependence on~$\ell$ from notation. Dimensional considerations necessitate that, indeed, there must be \textit{some} factor~$\lambda$, since a field line $\mathbf{f}$ is a curve in geometric space (with a dimension of length), while a vector field~$\mathbf{E}$ may be some dimensionally unrelated quantity (e.g. an electric field). \textit{At this point} we are only justified in assuming~$\lambda$ to be a constant. If $\lambda$ were allowed to vary with position---i.e. if~$\lambda$ were a scalar field $\lambda(\mathbf{r})$---then~(\ref{fl_start}) would give us a field line of a modified vector field \mbox{$\mathbf{E}'(\mathbf{r})=\lambda(\mathbf{r})\mathbf{E}(\mathbf{r})$}. We cannot assume \textit{in advance} that this would not affect~$\mathbf{f}$. However, if we were to write out~$\mathbf{f}$ and~$\mathbf{E}$ in some particular, e.g. Cartesian coordinates:
\begin{equation*}
\mathbf{f}=f_x\hat{\mathbf{x}}+f_y\hat{\mathbf{y}}+f_z\hat{\mathbf{z}} \quad\mbox{and}\quad \mathbf{E}=E_x\hat{\mathbf{x}}+E_y\hat{\mathbf{y}}+E_z\hat{\mathbf{z}},
\end{equation*}
then, by components, (\ref{fl_start}) reduces to:
\begin{equation}
\frac{\D f_x}{E_x}=\frac{\D f_y}{E_y}=\frac{\D f_z}{E_z} \quad(=\lambda\D\ell).
\end{equation}
Adopting one of the field line components as an independent variable---e.g.~$f_x$, making it a particular manifestation of a parameter~$\ell$---yields `operational' differential equations:
\begin{equation}
\frac{\D f_y}{\D f_x}=\frac{E_y}{E_x} \quad\mbox{and}\quad \frac{\D f_z}{\D f_x}=\frac{E_z}{E_x}
\label{operational}
\end{equation}
for the remaining components~$f_y$ and~$f_z$. These are often presented as self-evident forms of filed line equation(s), either in Cartesian or some other coordinates~\cite{electric_1,magnetic_1,dipole}. We see now that these equations---as a particular recasting of~(\ref{fl_start})---are (almost) completely independent of~$\lambda$. This means that we may, \textit{after all}, adopt for it an entire functional dependence~$\lambda(\mathbf{r})$, and it will not affect the sought field lines. Hence:
\begin{equation}
\frac{\D \mathbf{f}}{\D\ell}=\lambda(\mathbf{f})\mathbf{E}(\mathbf{f})
\label{fl_general}
\end{equation}
is \textit{the most general form of a field line equation}. We now have a freedom of choosing the best~$\lambda(\mathbf{r})$. There \textit{is} one requirement upon~$\lambda(\mathbf{r})$: it must be strictly positive, in order to preserve a vectorial direction of a modified field \mbox{$\lambda(\mathbf{r})\mathbf{E}(\mathbf{r})$}, wherever a direction of the original field $\mathbf{E}(\mathbf{r})$ is well defined\footnote{
Otherwise, a field line would be unjustifiably cut off at a point where \mbox{$\lambda(\mathbf{r})=\mathbf{0}$} and \mbox{$\mathbf{E}(\mathbf{r})\neq\mathbf{0}$}. If~$\lambda(\mathbf{r})$ were to change sign---either continuously or discontinuously---a `flow' of a  field line would change direction and it would fold in on itself. Thus,~$\lambda(\mathbf{r})$ can not be \textit{completely} arbitrary.
}. Thus:
\begin{equation*}
\lambda(\mathbf{r})>0 \quad\mbox{where}\quad \mathbf{E}(\mathbf{r})\neq\textbf{0}.
\end{equation*}
A particular selection now suggests itself, directly related to a vector field norm: \mbox{$\lambda(\mathbf{r})=1/|\mathbf{E}(\mathbf{r})|$}. Thus, the `best'  form of a field line equation~(\ref{fl_general}) becomes:
\begin{equation}
\frac{\D \mathbf{f}}{\D\ell}=\hat{\mathbf{E}}(\mathbf{f})=
\left\{\begin{array}{ccc}
\mathbf{E}(\mathbf{f})/|\mathbf{E}(\mathbf{f})| &\mbox{if}& \mathbf{E}(\mathbf{f})\neq\textbf{0},\\
\mathbf{0}&\mbox{if}& \mathbf{E}(\mathbf{f})=\textbf{0}.
\end{array}\right.
\label{fl_best}
\end{equation}
There are several advantages to this form\footnote{
Interestingly, \cite{magnetic_4,plotting} give a field line equation in a form:
\begin{equation*}
\mathbf{E}(\mathbf{f})\times\D\mathbf{f}=\mathbf{0},
\end{equation*}
since, by components,  it provides a same set of `operational' equations as in~(\ref{operational}). However, a familiar cross product is only defined in three spatial dimensions, so this form can not be generalized to higher dimensions. Thus,~(\ref{fl_best}) remains a superior form of a field line equation.
}. One is a clear and immediate interpretation of a parameter~$\ell$. Since a vector field is now a field of \textit{unit} magnitude (up to exceptional~$\textbf{0}$ values which we will consider understood from now on), a left hand side must also be of unit magnitude. This means that an increment~$\D\ell$ is equal to a length of a field line increment~$\D \mathbf{f}$. Thus,~$\ell$ itself is a cumulative \textit{arc length} of a given field line segment starting from some initial point~$\mathbf{f}_0$. The other advantage is a numerical stability in solving~(\ref{fl_best}) by numerical methods. This is especially important for an electric field of isolated charges, which varies significantly with a distance, thus possibly introducing numerical instabilities if uniform numerical values for a finite~$\Delta\ell$ were used. Finally, expressed using a unit field~$\hat{\mathbf{E}}$, (\ref{fl_best})~shows beyond any doubt that the field lines have no intrinsic connection to a magnitude of a vector field. In particular, unless some very specific conditions are met~\cite{redundant,electric_3}, a density of field lines does not provide any information about a field magnitude. A claim to the contrary is a widespread misconception that persists to this day.

%\clearpage

\end{document}